%% file: ZR.tex
\documentclass[sigconf]{acmart}

\usepackage{geometry}
\usepackage{balance}
\usepackage{bm}
\usepackage{threeparttable}
\usepackage{booktabs} 
\usepackage{multirow}
\usepackage{subfigure}
\usepackage{enumitem}
\usepackage{balance}
\usepackage{diagbox}
\usepackage[table]{xcolor}
\usepackage{comment}
\usepackage{enumitem}
\setlist[itemize]{leftmargin=*}
\usepackage{algorithm}
\usepackage{algpseudocode}

\usepackage{amsmath} 
 
\usepackage{amssymb}
\usepackage{pdfrender}

\usepackage[hang,flushmargin]{footmisc}

\newcommand{\cb}[1]{{\color{cyan}[cb: #1]}}
\newcommand{\dk}[1]{{\color{blue}[dk: #1]}}

\usepackage{etoolbox}
\AtBeginDocument{%
  }

\setcopyright{acmlicensed}
\copyrightyear{2018}
\acmYear{2018}
\acmDOI{XXXXXXX.XXXXXXX}
\acmConference[Conference acronym 'XX]{Make sure to enter the correct
  conference title from your rights confirmation emai}{June 03--05,
  2018}{Woodstock, NY}
\acmISBN{978-1-4503-XXXX-X/18/06}
\setcopyright{none}
\begin{document}

\title{SITA: Semantic Interest Tokens for Target-Aware Compression in Long-Sequence Recommendation}

\author{Rui Zhou}
\email{zhou_rui@mail.ustc.edu.cn}
\affiliation{%
  \institution{University of Science and Technology of China}
  \city{Hefei}
  \country{China}
}

\author{Bo Chen}
\email{renze03@kuaishou.com}
\affiliation{%
  \institution{Kuaishou Technology}
  \city{Beijing}
  \country{China}
}

\author{Qinglin Jia}
\email{dukang05@kuaishou.com}
\affiliation{
  \institution{Kuaishou Technology}
  \city{Beijing}
  \country{China}
}

\author{Jiezhou Ji}
\email{jijiezhou@kuaishou.com}
\affiliation{
  \institution{Kuaishou Technology}
  \city{Beijing}
  \country{China}
}

\author{Chaoyi Ma}
\email{machaoyi03@kuaishou.com}
\affiliation{%
  \institution{Kuaishou Technology}
  \city{Beijing}
  \country{China}
}

\author{Ruiming Tang}
\email{tangruiming@kuaishou.com}
\affiliation{%
  \institution{Kuaishou Technology}
  \city{Beijing}
  \country{China}
}

\author{Hao Wang}
\authornote{Corresponding author.}
\email{wanghao3@ustc.edu.cn;}
\affiliation{%
  \institution{University of Science and Technology of China}
  \city{Hefei}
  \country{China}
}

\author{Enhong Chen}
\email{cheneh@ustc.edu.cn}
\affiliation{%
  \institution{University of Science and Technology of China}
  \city{Hefei}
  \country{China}
}

\renewcommand{\shortauthors}{Rui Zhou et al.}
\renewcommand{\abstractname}{\uppercase{Abstract}} 
\renewcommand{\keywordsname}{\MakeUppercase{Keywords}}  

\begin{abstract}
    As user behavior histories continue to grow on modern Internet platforms, effectively modeling long behavior sequences has become crucial for predicting user interests in candidate items. Existing methods have evolved along two directions. One line dynamically retrieves target-relevant behaviors from long histories, enabling target-aware modeling but requiring target-dependent computation during inference. The other line compresses entire behavior sequences into compact user representations, achieving high efficiency and scalability but sacrificing target-specific adaptation due to target-independent encoding. 
    The key challenge is therefore to enable target-aware modeling while preserving the efficiency and scalability of compressed user representations.
    To address this challenge, we propose \textbf{SITA}, a target-aware compression framework for long-sequence recommendation. SITA enables target-aware compression by organizing compressed interests into semantic structures through semantic identifiers learned via parallel semantic quantization. Conditioned on the semantic identifier of the target item, SITA adaptively aggregates the corresponding structured interests to construct the target-specific user representation.
    Extensive experiments on public datasets and a large-scale industrial dataset demonstrate that SITA consistently outperforms representative baselines while maintaining strong scalability, highlighting its strong potential for real-world recommender systems.
\end{abstract}

\renewcommand{\arraystretch}{1.2}
\keywords{Long-Sequence Recommendation, Target-Aware Modeling, Interest Compression}

\maketitle

\input{sections/introduction.tex}

\input{sections/related_work.tex}

\input{sections/preliminary.tex}
\input{sections/method.tex}

\input{sections/experiments.tex}
\input{sections/conclusion.tex}
\balance

\newpage
\bibliographystyle{ACM-Reference-Format}
\bibliography{sample-base}

\end{document}

%% file: sections/introduction.tex
\section{INTRODUCTION}

Modern recommender systems provide personalized services by recommending items that best match users interests~\cite{hou2024cross,chen2021enhancing,wang2021dcn,wang2025dlf,zhou2025multi}. Historical interactions serve as the primary evidence for understanding users interests, as they contain rich behavioral signals that reflect users long-term and diverse preferences~\cite{zhang2026diet,ye2026fuxi,xie2025breaking,zhang2025killing,xie2024breaking}. Long-sequence modeling aims to capture holistic user interests from these historical interactions for accurate prediction over candidate items and ultimately improving recommendation performance~\cite{guo2021embedding, qin2020user,chen2026uniformer,zhou2024encode,gao2024hierrec}.

\begin{figure}[t]
  \centering
  \includegraphics[width=0.95\linewidth]{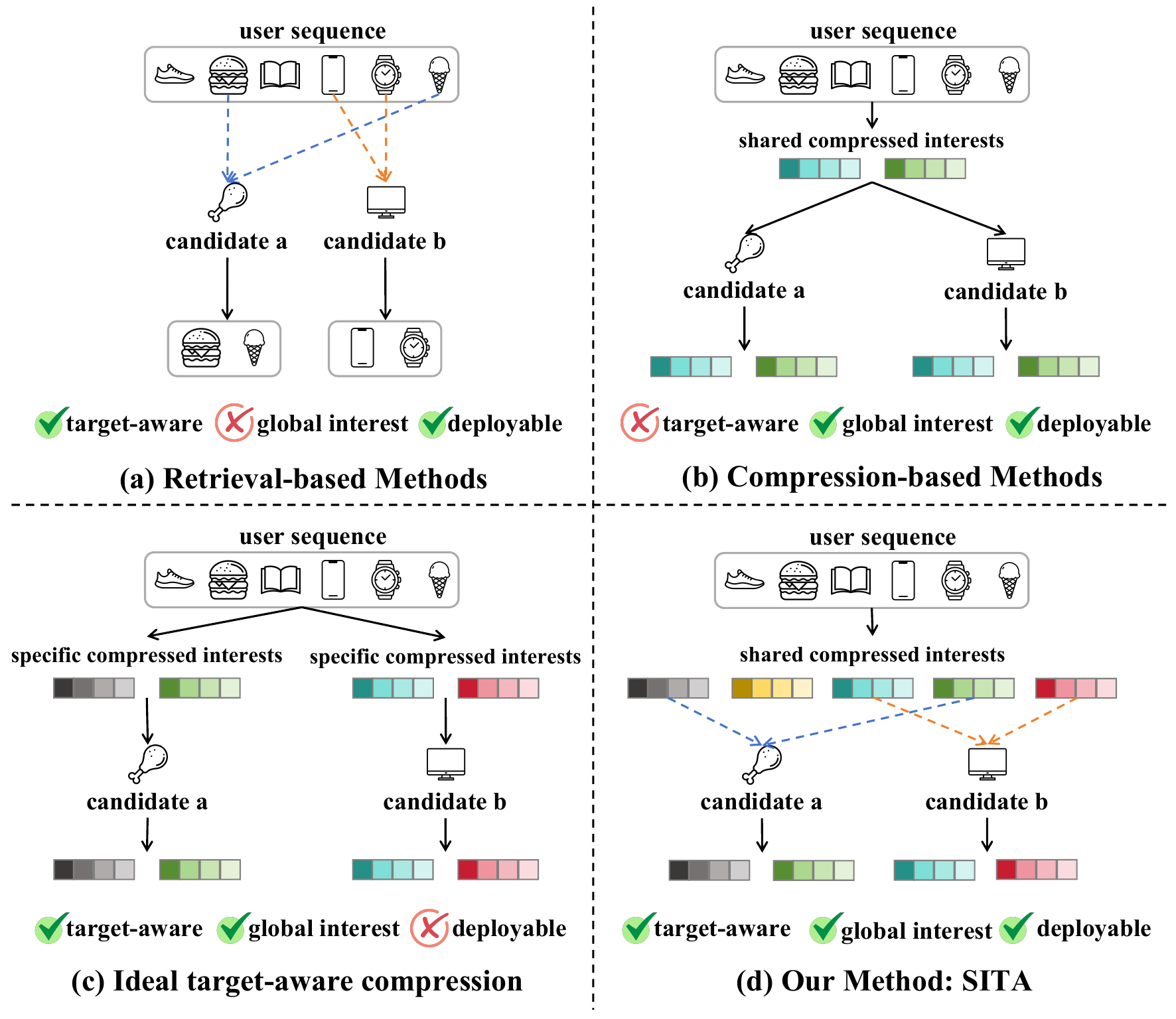}
  \caption{ Comparison of representative long-sequence modeling paradigms in terms of target-aware modeling, global user interest modeling, and real-world deployability. SITA is the only method that simultaneously satisfies all three desirable properties.}
  \label{fig:intro_tradeoff}
  \vspace{-6mm}
\end{figure}

To model long behavior sequences efficiently and effectively, existing methods have evolved along two representative paradigms. One line of work, represented by SIM~\cite{pi2020search} and TWIN~\cite{chang2023twin}, first efficiently retrieves the historical interactions most relevant to the target item from the original user behavior sequence and then effectively models user interests over the retrieved subsequence. These methods are referred to as retrieval-based methods, and the overall pipeline is illustrated in Figure~\ref{fig:intro_tradeoff}(a). By explicitly conditioning on the target item during retrieval, they effectively capture highly target-specific user interests. However, filtering the behavior sequence inevitably discards a large portion of historical interactions, limiting the model to only local user interests at the expense of global user interests.

Another line of work includes C-Former~\cite{wang2025transformers} and VISTA~\cite{chen2025massive}, first compresses the complete behavior sequence into a compact set of reusable interest representations and then performs user interest modeling over these compressed representations. These methods are referred to as compression-based methods, as illustrated in Figure~\ref{fig:intro_tradeoff}(b). By modeling the complete behavior history and decoupling behavior sequence encoding from target-specific interaction, they effectively capture global user interests while enabling efficient online inference. However, under this design, all target items have to share the same reusable interest representations for each user, making the learned user interests inherently target-agnostic.

The complementary strengths and limitations of these two representative paradigms together reveal a fundamental trade-off between target-aware modeling and global interest modeling. A natural solution is to maintain target-specific compressed user interests for each target item, as illustrated in Figure~\ref{fig:intro_tradeoff}(c), thereby enabling both target-aware modeling and global user interest modeling. 
However, such a solution is not deployable in real-world recommender systems, since maintaining dedicated compressed interest representations for every user--item pair incurs an $O(|\mathcal{U}||\mathcal{V}|)$ storage complexity, which is prohibitively expensive in practice.
These observations give rise to two fundamental challenges for long-sequence modeling:
(1) \textbf{Ensuring Deployability}. How to realize such a modeling paradigm at the massive user--item scale without incurring prohibitive storage and maintenance costs.
(2) \textbf{Achieving Target Awareness and Global Interest Modeling}. How to construct compressed user interests that preserve global behavioral information while supporting target-aware user interest modeling.


To address these challenges, we propose \textbf{SITA}, as illustrated in Figure~\ref{fig:intro_tradeoff}(d), a novel long-sequence modeling paradigm that simultaneously achieves real-world deployability and target-aware global interest modeling. Specifically, SITA first performs semantic-based item encoding through Balanced Parallel Quantization (BPQ), transforming the original item space into a compact semantic codebook space of fixed size $M$, thereby reducing the storage complexity from $O(|\mathcal{U}||\mathcal{V}|)$ to $O(M|\mathcal{U}|)$ and making the proposed paradigm readily deployable at industrial scale.
Based on this semantic organization, SITA employs stacked Structured Interest Compression (SIC) blocks to compress the complete behavior sequence into a set of structured interest tokens, where \textit{intra-group modeling} captures fine-grained user interests within each semantic group and \textit{inter-group interaction} enables complementary information exchange across semantic groups. The resulting interest tokens are stored as user representations that preserve global user interests while being organized according to the semantic space, thereby enabling target-aware selection. During inference, the SID-Guided Selection (SGS) module selects the corresponding interest tokens according to the candidate item's semantic identifiers to produce the final target-aware user interest representation.
Our main contributions are summarized as follows:
\begin{itemize}
    \item We revisit existing long-sequence modeling paradigms, comprehensively analyze the fundamental trade-off between target-aware modeling and global user interest modeling, and propose a novel target-aware compression paradigm that simultaneously achieves efficient real-world deployability and target-aware global interest modeling.
    \item We propose \textbf{SITA}, a novel long-sequence modeling paradigm that simultaneously achieves real-world deployability and target-aware global interest modeling. Specifically, SITA performs semantic encoding to transform the original item space into a compact semantic space for deployment, and learns semantically organized interest tokens through compression blocks with group-level interest modeling, preserving global user interests while enabling target-aware selection via semantic identifiers.
    \item Extensive experiments on both public datasets and a large-scale industrial recommender system demonstrate that SITA effectively achieves the proposed design objective, consistently outperforming existing long-sequence recommendation methods while maintaining high efficiency and real-world deployability.
\end{itemize}

%% file: sections/related_work.tex
\section{RELATED WORK}

\subsection{Long-Sequence Modeling}

Long-sequence modeling aims to capture informative user interests from extensive historical interactions for accurate prediction over target items~\cite{pan2025revisiting,ye2025fuxi,zhang2025td3,shen2026p}. As user behavior histories continue to grow, numerous methods have been proposed to model increasingly long behavior sequences, gradually evolving into two representative paradigms: retrieval-based methods and compression-based methods~\cite{zhou2026survey}.

\textit{Retrieval-based methods} first retrieve the historical interactions most relevant to the target item from the original behavior sequence and then perform user interest modeling over the retrieved subsequence. SIM~\cite{pi2020search} first introduced this paradigm and achieved promising performance. ETA~\cite{chen2021end} and MIRRN~\cite{xu2025multi} further improved retrieval efficiency through hash-based indexing and Hamming-distance. TWIN~\cite{chang2023twin} and DARE~\cite{feng2025long} addressed the inconsistency issues in retrieval-based modeling. More recent works, including LIC~\cite{zhu2025long}, MUSE~\cite{wu2025muse}, and RAL-CDNet~\cite{tang2025retrieval}, further enhanced retrieval-based modeling by incorporating temporal information, cross-domain knowledge, and multimodal information, respectively.
These methods improve retrieval quality and efficiency from different perspectives while following the same retrieval-based paradigm.

\textit{Compression-based methods} first compress the complete behavior sequence into a compact set of shared interest representations and then perform downstream interest modeling over the compressed representations. Early works, such as MIMN~\cite{pi2019practice} and HPMN~\cite{ren2019lifelong}, employed memory networks for sequence compression, while ENCODE~\cite{zhou2024encode} and TWIN V2~\cite{si2024twin} adopted clustering-based approaches. More recent methods, including C-Former~\cite{wang2025transformers} and VISTA~\cite{chen2025massive}, perform target-aware interaction over compressed interests. Other approaches, such as Trinity~\cite{yan2024trinity}, CHIME~\cite{bai2025chime}, and DMQN~\cite{wei2025deep}, further improve compression quality through histogram-based representations and semantic codebooks.
These methods mainly differ in how compressed interests are constructed while sharing the same compression-based paradigm. However, their shared compressed representations limit target-aware interest modeling.
Our work revisits this paradigm to achieve target-aware global interest modeling under real-world deployment constraints.


\subsection{Semantic Identifiers}

Item identifiers are the foundation of modern recommender systems and have recently evolved from random discrete IDs to semantic identifiers (SIDs), which encode semantic similarity among items in compact discrete representations~\cite{luo2025qarm,li2026sid}. Existing SID methods are primarily built upon residual quantization, including RQ-KMeans~\cite{deng2025onerec}, which performs residual clustering, and RQ-VAE~\cite{shan2025automatic}, which learns semantic identifiers through end-to-end neural quantization with reconstruction and commitment objectives.

Building upon this foundation, recent studies have further improved SID learning from the perspectives of quantization strategies, codebook architectures, and optimization objectives. For example, COINS~\cite{zhao2026coins}, QARM V2~\cite{xia2026qarm}, and SA$^2$CRQ~\cite{wang2026towards} enhance semantic representations through hybrid or variable-length semantic identifiers. STORE~\cite{xu2025store} constructs parallel codebooks using a mixture-of-experts architecture, while QuaSID~\cite{hu2026stop} introduces conflict-aware regularization to alleviate code collisions.

%% file: sections/preliminary.tex
\section{PRELIMINARY}

In modern recommender systems, the long-sequence modeling module aims to capture user interests from historical behaviors for downstream ranking. Let $\mathcal{U}$ and $\mathcal{V}$ denote the sets of users and items, respectively. For a user $u\in\mathcal{U}$, the historical behavior sequence is denoted as $\mathcal{S}_{u}=[v_1,v_2,\ldots,v_L]$, where $v_i\in\mathcal{V}$ and $L$ is the sequence length.

Given a target item $v_t\in\mathcal{V}$, user profile features $\mathbf{x}^{u}$, and contextual features $\mathbf{x}^{c}$, an ideal long-sequence modeling module is expected to learn a target-aware global user interest:
\begin{equation}
    \mathbf{h}_{u,t}=f(\mathcal{S}_u,v_t,\mathbf{x}^{u},\mathbf{x}^{c}),
\end{equation}
where the representation dynamically captures the user's global interests relevant to the target item.

However, existing compression-based long-sequence modeling methods usually decouple sequence compression from target items. Specifically, they first compress the entire behavior sequence into a fixed user interest:
\begin{equation}
    \mathbf{h}_{u}=f(\mathcal{S}_u,\mathbf{x}^{u},\mathbf{x}^{c}),
\end{equation}
which summarizes the user's global interests independently of different target items. Although such methods significantly reduce the computational cost for long sequences, the target-independent representation $\mathbf{h}_{u}$ cannot adaptively emphasize different interest aspects according to the target item.

Therefore, a key challenge is to bridge the gap between efficient sequence compression and target-aware interest modeling, enabling the learned interest representation to adapt to different target items without introducing prohibitive computational costs.

%% file: sections/method.tex
\begin{figure*}[t]
  \centering
  \includegraphics[width=0.95\linewidth]{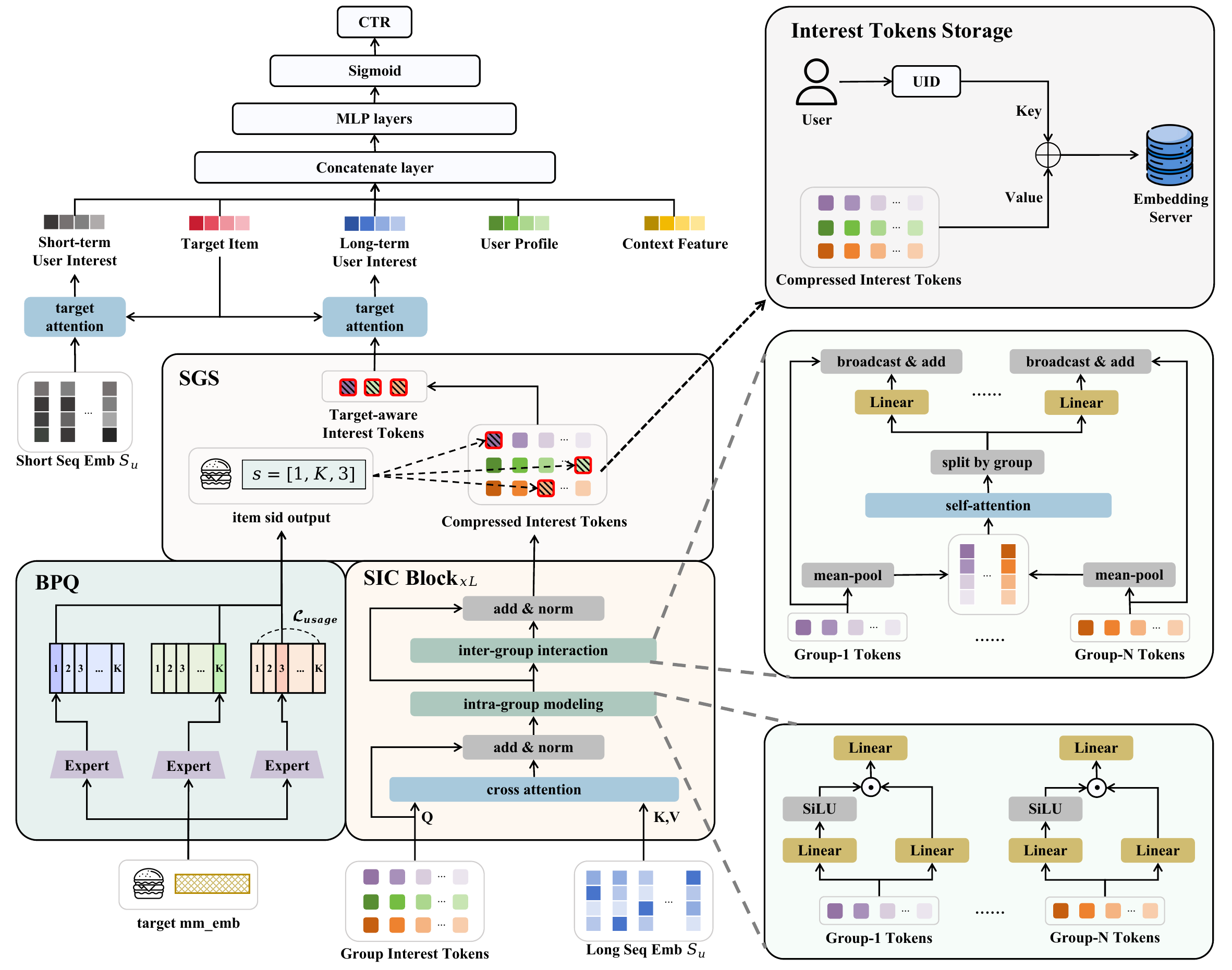}
  \caption{Overview of SITA. BPQ maps each item into a semantic identifier (SID). SIC compresses the original behavior sequence into semantically organized interest tokens through intra-group refinement and inter-group communication, where the resulting interest tokens are stored as compact user representations. During inference, SGS selects the corresponding interest tokens according to the candidate item's SID, and a target attention module aggregates the selected tokens to produce the final target-aware long-term user interest representation, which is concatenated with other features for the ranking model.}
  \label{fig:SITA}
\end{figure*}

\section{METHODOLOGY}
In this section, we first introduce the overall architecture of SITA. We then describe its three key components in details, namely Balanced Parallel Quantization (BPQ), Structured Interest Compression (SIC), and SID-Guided Selection (SGS). Finally, we analyze the complexity of SITA.

\subsection{Overall}
Figure~\ref{fig:SITA} illustrates the overall architecture of SITA, which consists of three core components: Balanced Parallel Quantization (BPQ), Structured Interest Compression (SIC), and SID-Guided Selection (SGS). To enable target-aware global interest modeling under real-world deployment constraints, SITA introduces a compact semantic space to semantically organize interest tokens, enabling candidate items to dynamically select different interest tokens according to their semantic identifiers.
Specifically, BPQ learns structured semantic identifiers from item-side multimodal representations. Guided by the semantic encoding structure, SIC compresses the original behavior sequence into semantically organized interest tokens through stacked compression blocks, which are stored as compact user interests. Given a target item, SGS selects a different subset of interest tokens according to its semantic identifier for different target items, thereby enabling target-aware global interest modeling. 
The following subsections describe these three components in detail.


\subsection{Balanced Parallel Quantization}
Modern recommender systems typically involve billions of items. Directly maintaining target-specific compressed interests for every user-item pair would require storage complexity of $O(|\mathcal{U}||\mathcal{V}|)$, which is prohibitive in practice. To make target-aware compression scalable, we first decompose the original item space into a compact structured encoding space, where a combinatorial number of item identities can be represented by composing a small set of encoding units. To further preserve semantic consistency among related items, the structured encoding space should capture multiple complementary semantic aspects, so that semantically similar items obtain similar structured encodings. To this end, we propose \textbf{Balanced Parallel Quantization} (BPQ), which learns such a structured semantic encoding space from item-side multimodal representations through multiple parallel semantic codebooks.


Formally, given an item multimodal embedding $\mathbf{x}\in\mathbb{R}^{d_m}$, where $d_m$ is the multimodal embedding dimension, we employ $N$ independent experts to project it into $N$ latent vectors:
\begin{equation}
\mathbf{e}_n = g_n(\mathbf{x}), \quad n=1,2,\dots,N,
\end{equation}
where each $g_n(\cdot)$ is implemented as a multi-layer perceptron (MLP). Each latent vector $\mathbf{e}_n$ is then quantized using codebook 
$\mathcal{C}_n=\{\mathbf{c}_{n,1},\mathbf{c}_{n,2},\dots,\mathbf{c}_{n,K}\}$,
which contains $K$ codewords. The nearest codeword index forms one component of the semantic identifier:
\begin{equation}
s_n=\arg\min_{k\in\{1,\dots,K\}}
\|\mathbf{e}_n-\mathbf{c}_{n,k}\|_2^2.
\end{equation}
Accordingly, each item is assigned a structured semantic identifier
\begin{equation}
\mathbf{s}=[s_1,s_2,\dots,s_N],\quad
s_n\in\{1,\dots,K\},
\end{equation}
where $N$ and $K$ denote the number of codebooks and the number of codewords per codebook, respectively.

The selected codewords from all codebooks are concatenated and fed into an MLP decoder to reconstruct the original multimodal representation. The reconstruction loss is computed using the mean squared error (MSE) between the reconstructed and original representations. Following the standard vector quantization formulation, the training objective also includes the codebook and commitment losses. 
Beyond these standard objectives, BPQ further introduces a usage-balance regularization to prevent codeword collapse. Specifically, we compute the average codeword usage distribution $\bar{\mathbf p}_n$ for each codebook by averaging the soft assignment probabilities over a mini-batch, and encourage it to approach the uniform distribution by minimizing
\begin{equation}
    \mathcal{L}_{\mathrm{usage}} = \frac{1}{N}\sum_{n=1}^{N}\left\|\bar{\mathbf{p}}_n-\frac{1}{K}\mathbf{1}\right\|_2^2.
\end{equation} 



After training, BPQ produces a structured semantic space together with a fixed semantic identifier for each item. By composing semantic codes from $N$ parallel codebooks, each containing $K$ codewords, BPQ represents up to $K^N$ semantic identifiers using only $N×K$ encoding units, reducing the storage complexity from $O(|\mathcal{U}||\mathcal{V}|)$ to $O(|\mathcal{U}|NK)$. This design objective fundamentally differs from that of conventional quantization methods, such as residual quantization, which are primarily developed to improve reconstruction fidelity by progressively encoding residuals. Instead, BPQ is designed to construct a compositional encoding space, where multiple parallel semantic codebooks capture complementary semantic aspects and jointly form structured semantic identifiers. These semantic identifiers are subsequently leveraged by SGS to select target-specific interest tokens for target-aware global interest modeling.

\subsection{Structured Interest Compression
Block}

After BPQ learns a compact semantic space, \textbf{Structured Interest Compression} (SIC) blocks compress the original behavior sequence into a set of semantically organized interest tokens. Specifically, we initialize a set of learnable interest tokens:
\begin{equation}
    \mathbf{Z}_u^{(0)}\in\mathbb{R}^{NK\times d},
\end{equation}
where $N$ denotes the number of semantic groups, $K$ is the number of interest tokens in each group, and $d$ is the hidden dimension.


Starting from the initialized interest tokens, SIC progressively refines their representations through a stack of compression blocks. Within the $l$-th compression block, the user behavior sequence is first incorporated into the interest tokens via a cross-attention layer, yielding the updated interest representation $\mathbf{H}_u^{(l)}$:
\begin{equation}
\mathbf{H}_u^{(l)}
=
\operatorname{CrossAtt}^{(l)}
\left(
\mathbf{Z}_u^{(l-1)},
\mathbf{S}_u, \mathbf{S}_u
\right)
+
\mathbf{Z}_u^{(l-1)},
\end{equation}
where $\mathbf{Z}_u^{(l-1)}$ denotes the interest tokens from the $(l-1)$-th compression block, $\mathbf{S}_u$ is the embedded user behavior sequence. 
The updated interest tokens are subsequently refined through group-level modeling. To align the learned interests with the structured encoding constructed by BPQ, the interest tokens are organized into $N$ groups, each containing $K$ interest tokens. We then perform intra-group modeling and inter-group interaction to progressively learn structured user interests. After stacking multiple SIC blocks, the resulting structured interest tokens are stored as a compact user-specific token set for subsequent target-aware selection. The following subsections describe these three components: \textit{intra-group modeling}, \textit{inter-group interaction}, and \textit{interest token storage}.

\subsubsection{Intra-group Modeling}



After interest extraction, the interest tokens have incorporated user behavior information through cross-attention. Since the interest tokens are semantically organized into different groups, sharing a single feed-forward network across all groups may limit their ability to learn group-specific representations. Therefore, we assign per-group feed-forward network to each semantic group. Each feed-forward network is implemented as a SwiGLU block for improved expressive capacity:

\begin{equation}
\begin{aligned}
\mathbf{U}_{u,n}^{(l)}
&=
\operatorname{SiLU}
\!\left(
\mathbf{H}_{u,n}^{(l)}
\mathbf{W}_{n}^{(g)}
\right)
\odot
\left(
\mathbf{H}_{u,n}^{(l)}
\mathbf{W}_{n}^{(u)}
\right),
\\
\mathbf{G}_{u,n}^{(l)}
&=
\mathbf{U}_{u,n}^{(l)}
\mathbf{W}_{n}^{(o)},
\end{aligned}
\end{equation}
where
$\mathbf{H}_{u,n}^{(l)}\in\mathbb{R}^{K\times d}$ denotes the $n$-th group of interest tokens after interest extraction, and
$\{\mathbf{W}_{n}^{(g)},\mathbf{W}_{n}^{(u)}, \mathbf{W}_{n}^{(o)}\}$
are the learnable parameters of the group-specific SwiGLU block for the corresponding semantic group.

Compared with a shared feed-forward network, the proposed group-specific transformation allows each semantic group to learn distinct nonlinear transformations, leading to more expressive group-specific representations. The resulting interest tokens are then passed to the inter-group interaction stage to exchange complementary information across different semantic groups.

\subsubsection{Inter-group Interaction}


After intra-group modeling, each semantic group has learned its own group-specific representations. However, user interests are inherently correlated rather than completely independent, making interactions across different semantic groups necessary for capturing complementary information. A straightforward solution is to perform token-level self-attention over all interest tokens. However, such unrestricted token-level interactions undermine the semantic grouping of the interest tokens, which serves as the foundation for subsequent SID-guided activation. Therefore, instead of directly modeling interactions among all interest tokens, we first summarize each semantic group by mean pooling, obtaining N group-level summaries. Self-attention is then performed over these summaries to capture interactions among different semantic groups, producing the updated group representation $\mathbf{R}_{u}^{(l)}$:
\begin{equation}
\mathbf{R}_{u}^{(l)}
=
\operatorname{SelfAtt}^{(l)}
\left(
\left[
\frac{1}{K}\sum_{i=1}^{K}\mathbf{G}_{u,1,i}^{(l)},
\ldots,
\frac{1}{K}\sum_{i=1}^{K}\mathbf{G}_{u,N,i}^{(l)}
\right]
\right).
\end{equation}

The refined group representations are transformed into group-specific bias vectors and broadcast to all interest tokens within the corresponding semantic group:
\begin{equation}
\mathbf{Z}_{u,n}^{(l)}
\leftarrow
\mathbf{G}_{u,n}^{(l)}
+
\operatorname{MLP}
\left(
\mathbf{R}_{u,n}^{(l)}
\right),
\end{equation}
where $\mathbf{Z}_{u,n}^{(l)}$ denotes the updated interest tokens of the n-th semantic group after incorporating the group-specific bias. The updated tokens from all $N$ semantic groups are concatenated to form $\mathbf{Z}_u^{(l)}\in\mathbb{R}^{NK\times d}$ , which serves as the output of the $l$-th SIC block. By performing interaction at the group level rather than the token level, the proposed module preserves the semantic grouping of the interest tokens while enabling information exchange across different semantic groups. In addition, the interaction complexity is reduced from $O((NK)^2)$ to $O(N^2)$, making the proposed design more efficient for large-scale recommendation systems. 

\subsubsection{Interest tokens storage.}
After stacking $L$ SIC blocks, the refined interest tokens $\mathbf{Z}_{u}^{(L)}$ are obtained for each user.Since the entire SIC process is target-agnostic, $\mathbf{Z}_{u}^{(L)}$
can be computed offline and stored as a reusable set of user-specific interest tokens. During online inference, the stored interest tokens can be directly accessed without repeatedly interacting with the original behavior sequence, thereby decoupling online inference from long-sequence encoding. Consequently, only $N\times K$ interest tokens are maintained for each user, reducing the storage complexity from $O(|\mathcal{U}||\mathcal{V}|)$ to $O(|\mathcal{U}|NK)$. Together with the compositional semantic encoding space constructed by BPQ, these $N\times K$ stored interest tokens support up to $K^N$ target-specific selection patterns. Since $N$ and $K$ are configurable, $N\times K$ can be kept on the order of hundreds in practice while still providing a combinatorially large semantic encoding space, making target-aware compression practically deployable at industrial scale. In the subsequent SGS module, the semantic identifier of the target item is used to select the corresponding subset of interest tokens from $\mathbf{Z}_{u}^{(L)}$ for target-aware global interest modeling.

\begin{table}
\centering
\caption{Inference complexity comparison of representative long-sequence modeling methods.}
\label{tab:complexity}
\renewcommand{\arraystretch}{0.95}
\begin{tabular}{lc}
\toprule
\textbf{Model} & \textbf{Inference Complexity} \\
\midrule
SIM-Hard & $O(B\log A + BRd)$ \\
SIM-Soft & $O(BLd + BRd)$ \\
ETA & $O(BLm + BRd)$ \\
TWIN & $O(BLC+BRd)$ \\
C-Former & $O(BTd)$ \\
VISTA & $O(BTd)$ \\ \hline
\textbf{SITA} & $O(BNd)$ \\
\bottomrule
\end{tabular}

\vspace{1mm}
\footnotesize
$B$: number of candidate items processed simultaneously during inference;
$d$: embedding dimension;
$L$: user sequence length;
$R$: retrieved sequence length;
$T$: number of compressed interest tokens;
$N$: number of semantic groups;
$A$: size of the attribute inverted index in SIM-Hard;
$m$: number of hash functions in ETA;
$C$: number of user-item cross features in TWIN.

\end{table}

\begin{table}
  \centering
  \caption{Statistics of the datasets.}
  \label{tab:dataset_statistics}
  \renewcommand{\arraystretch}{0.95}
  \setlength{\tabcolsep}{5.5pt}
  \begin{tabular}{lrrrrr}
    \toprule
    Dataset & Fields & Users & Items & Samples & Length \\
    \midrule
    Taobao-MM & 13 & 8.79M & 35.40M & 99.00M & 1,000 \\
    XLong     & 4  & 20.00K    & 3.27M  & 0.14M    & 1,000 \\
    \bottomrule
  \end{tabular}
\end{table}

\subsection{SID-Guided Selection}

\begin{table*}[t]
  \centering
  \caption{Performance comparison on the Taobao-MM and XLong datasets. The best and second-best results are highlighted in bold and underlined, respectively. $^\dagger$ denotes that SITA significantly outperforms the strongest baseline according to a paired t-test with $p<0.05$.}
  \label{tab:main_results}
  \setlength{\tabcolsep}{8pt}
  \renewcommand{\arraystretch}{0.95}
  \begin{tabular*}{0.95\textwidth}{@{\extracolsep{\fill}}lc>{\columncolor{gray!15}}c c>{\columncolor{gray!15}}c c>{\columncolor{gray!15}}c c>{\columncolor{gray!15}}c}
    \toprule
    \multirow{2}{*}{Model} & \multicolumn{4}{c}{Taobao-MM} & \multicolumn{4}{c}{XLong} \\
    \cmidrule(lr){2-5} \cmidrule(lr){6-9}
    & AUC & Impr. & GAUC & Impr. & AUC & Impr. & GAUC & Impr. \\
    \midrule
    DIN      & 0.6358 & --    & 0.6081 & --    & 0.8817 & --    & 0.8785 & --    \\
    SIM & 0.6428 & 1.10\% & 0.6130 & 0.82\% & 0.8824 & 0.09\% & 0.8758 & -0.30\% \\
    TWIN     & 0.6452 & 1.47\% & 0.6099 & 0.30\% & 0.8851 & 0.39\% & 0.8811 & 0.30\% \\
    MUSE     & 0.6447 & 1.40\% & 0.6149 & 1.12\% & 0.9008 & 2.17\% & 0.8984 & 2.27\% \\
    C-Former & \underline{0.6543} & \underline{2.91\%} & \underline{0.6162} & \underline{1.34\%} & 0.8934 & 1.32\% & 0.8889 & 1.19\% \\
    UxSID    & 0.6542 & 2.89\% & 0.6159 & 1.29\% & 0.9006 & 2.15\% & 0.8971 & 2.12\% \\
    STCA     & 0.6531 & 2.72\% & 0.6151 & 1.16\% & 0.8939 & 1.38\% & 0.8894 & 1.24\% \\
    LONGER   & 0.6514 & 2.45\% & 0.6138 & 0.95\% & \underline{0.9036} & \underline{2.48\%} & \underline{0.9003} & \underline{2.48\%} \\
    \midrule
    SITA  & \textbf{0.6550} $\dagger$ & \textbf{3.02\%} & \textbf{0.6175} $\dagger$ & \textbf{1.56\%} & \textbf{0.9149} $\dagger$ & \textbf{3.77\%} & \textbf{0.9120} $\dagger$ & \textbf{3.82\%} \\
    \bottomrule
  \end{tabular*}
\end{table*}


After SIC, SITA obtains a compact set of semantically organized interest tokens $\mathbf{Z}_{u}^{(L)}$, where each semantic group contains $K$ interest tokens corresponding to the $K$ semantic identifiers learned by BPQ. Given the stored interest tokens together with the semantic identifier of a target item, SGS constructs a target-aware interest token set by selecting the corresponding interest token from each semantic group. Consequently, different target items select different subsets of interest tokens, enabling target-aware global
interest modeling while reusing the same compact user-specific interest tokens.

Specifically, let the semantic identifier of the target item be
\begin{equation}
\label{eq:sid_seq}
\mathbf{s}_t=(s_1,s_2,\ldots,s_N),
\end{equation}
where $s_n\in\{1,\ldots,K\}$ denotes the semantic identifier assigned by the
$n$-th semantic codebook. For each semantic group, the corresponding interest
token is selected as
\begin{equation}
\mathbf{e}_{u,n}
=
\mathbf{Z}_{u,n,s_n},
\end{equation}
where $\mathbf{Z}_{u,n,s_n}$ denotes the $s_n$-th interest token in the
$n$-th semantic group. The selected interest tokens are then collected as
\begin{equation}
\mathbf{E}_{u,t}
=
\left[
\mathbf{e}_{u,1},
\mathbf{e}_{u,2},
\ldots,
\mathbf{e}_{u,N}
\right]
\in
\mathbb{R}^{N\times d}.
\end{equation}

The resulting $\mathbf{E}_{u,t}$ constitutes the target-aware global interest of user $u$ with respect to the target item $t$, which is subsequently fed into the subsequent recommendation model for final prediction.



\subsection{Complexity Analysis}


We analyze the online inference complexity of SITA. Since modern recommender systems typically score a large number of candidate items for each user, inference efficiency is critical for large-scale online serving~\cite{xia2025transact,wei2026dmgin}.

During online inference, given a target item, SITA first performs SID-guided token selection to obtain the target-aware interest tokens. Let $B$ denote the number of candidate items processed simultaneously, $N$ the number of semantic groups, and $d$ the embedding dimension. This selection requires one index lookup in each semantic group, resulting in a complexity of $O(BN)$. The subsequent target attention over the N selected interest tokens has a complexity of $O(BNd)$. Since $d \gg 1$, the index lookup cost is negligible compared with the attention computation. Therefore, the overall online inference complexity of SITA is $O(BNd)$.

Table~\ref{tab:complexity} compares the online inference complexity of SITA with representative long-sequence recommendation methods. SITA maintains competitive online inference complexity while eliminating interactions between the target item and the original behavior sequence during online inference. Compared with existing compression-based methods, SITA preserves the same order of online inference complexity while enabling target-aware global interest modeling instead of relying on shared user interests.

%% file: sections/experiments.tex
\section{EXPERIMENTS}
\subsection{Experimental Settings}
\subsubsection{Datasets.}
We evaluate SITA on two real-world public datasets, \textbf{Taobao-MM}~\cite{wu2025muse} and \textbf{XLong}~\cite{ren2019lifelong}, which comprise features from multiple fields, including users' historical behavior sequences and multimodal representations. 
For a fair evaluation, the datasets are partitioned chronologically. The multimodal information is consistently used across all models, and only the long-sequence modeling modules are modified for experiments with different long-sequence modeling methods. Table~\ref{tab:dataset_statistics} summarizes the detailed statistics of the public datasets used in our experiments.

\subsubsection{Evaluation Protocols.}
We evaluate model performance using two metrics: \textbf{Area Under the ROC Curve (AUC)} and \textbf{Group Area Under the ROC Curve (GAUC)}~\cite{yin2025feature,wang2025enhancing}. AUC measures the probability that the model ranks a randomly selected positive instance higher than a randomly selected negative instance, reflecting its overall ranking capability. GAUC computes the AUC for each user (or group) separately and then aggregates the results using a weighted average based on the number of impressions, providing a more reliable evaluation of personalized recommendation performance.

\subsubsection{Baselines.}
To comprehensively evaluate the effectiveness of SITA, we compare it with a diverse set of representative baselines spanning classical retrieval-based methods, compression-based long-sequence recommendation methods, and several recent state-of-the-art approaches. The detailed descriptions of these baselines are provided below.
\begin{itemize}
    \item \textbf{DIN}~\cite{zhou2018deep} is a classical target-aware sequential recommendation method that models user interests conditioned on the target item.
    
    \item \textbf{SIM}~\cite{pi2020search} is a representative retrieval-based long-sequence recommendation method that first retrieves target-relevant behaviors from the user history and then performs fine-grained interest modeling.
    
    \item \textbf{TWIN}~\cite{chang2023twin} is a retrieval-based method that improves the consistency between target-aware behavior retrieval and downstream interaction modeling for long user behavior sequences.
    
    \item \textbf{MUSE}~\cite{wu2025muse} is a multimodal retrieval-based method that leverages multimodal item representations to enhance long-term user interest modeling.
    
    \item \textbf{C-Former}~\cite{wang2025transformers} is a representative compression-based method that compresses long behavior sequences into a compact set of interest representations, which are shared across different target items for the same user.
    
    \item \textbf{UxSID}~\cite{zhang2026uxsid} is a compression-based method that learns semantic IDs for items and incorporates them as semantic features for user interest modeling.
    
     \item \textbf{STCA}~\cite{guan2026make} is a recent long-sequence recommendation method that models user behaviors through stacked target cross-attention.
    
    \item \textbf{LONGER}~\cite{chai2025longer} is a recent long-sequence recommendation method that combines token merging with hybrid attention for efficient long-sequence modeling.
\end{itemize}

\subsubsection{Implementation Details.}

For fair comparison, all models are trained under the same training protocol for the same epoch. Sparse and dense parameters are optimized using SparseAdam and AdamW, with learning rates of $2\times10^{-3}$ and $2\times10^{-4}$, respectively. The embedding dimension is fixed at 16 for all models. For our proposed SITA, we stack 4 SIC blocks. The number of parallel codebooks $N$ and the codebook size $K$ are set to $(16,16)$ on Taobao-MM and $(8,16)$ on XLong, respectively.

\subsection{Overall Performance}
We conduct comprehensive extensive experiments to compare SITA with representative long-sequence modeling methods, while using DIN as the common baseline for all comparisons.
The results are summarized in Table~\ref{tab:main_results}.
SITA consistently achieves the best performance across all evaluation metrics, from which we draw the following key observations.
\begin{itemize}
    \item \textbf{Necessity of long-sequence modeling.} Compared with the classical sequential recommendation baseline DIN, almost all long-sequence recommendation methods achieve consistent performance improvements on both datasets, demonstrating that dedicated long-sequence modeling is essential for accurately capturing user interests.
    
    \item \textbf{Advantage over retrieval-based methods.}
    Compared with retrieval-based methods such as TWIN, SITA achieves consistent improvements on both datasets. Retrieval-based methods capture target-aware interests by selecting relevant behaviors, but may overlook diverse user interests beyond the retrieved subsequence. By preserving global interests while enabling target-aware selection, SITA achieves more comprehensive interest modeling.

    \item \textbf{Advantage over compression-based methods.}
    Compared with compression-based methods such as C-Former, SITA demonstrates significant performance gains across both datasets. Existing compression-based methods efficiently capture global interests but produce target-independent representations, limiting their ability to model item-related preferences. SITA performs target-aware selection on structured interest tokens, enabling the extraction of item-relevant interests from compressed representations.

    \item \textbf{Advantage over recent sequence modeling methods.}
    Compared with recent sequence modeling methods such as LONGER, SITA further improves performance by combining expressive sequence modeling with target-aware global interest modeling. Specifically, the stacked SIC blocks perform intra-group modeling and inter-group interaction to learn more expressive and structured interest representations.
\end{itemize}


\subsection{Ablation Study}
We conduct ablation studies on the XLong dataset to evaluate the contribution of each key component in SITA. Specifically, we consider the following variants: 
\begin{itemize}
    \item \textbf{SITA w/o IGM.} Replacing the group-specific SwiGLU modules with a single shared SwiGLU across all semantic groups. 

    \item \textbf{SITA w/o IGI.} Removing the inter-group interaction module. 

    \item \textbf{SITA w/o SGS.} Removing SID-Guided Selection, causing all target items to share the same interest token set. 

    \item \textbf{SITA w/ SimGS.} Replacing SID-Guided Selection module with similarity-based Top-$N$ selection while keeping the number of selected interest tokens unchanged. 
\end{itemize}

\begin{figure}
  \centering
  \includegraphics[width=\linewidth]{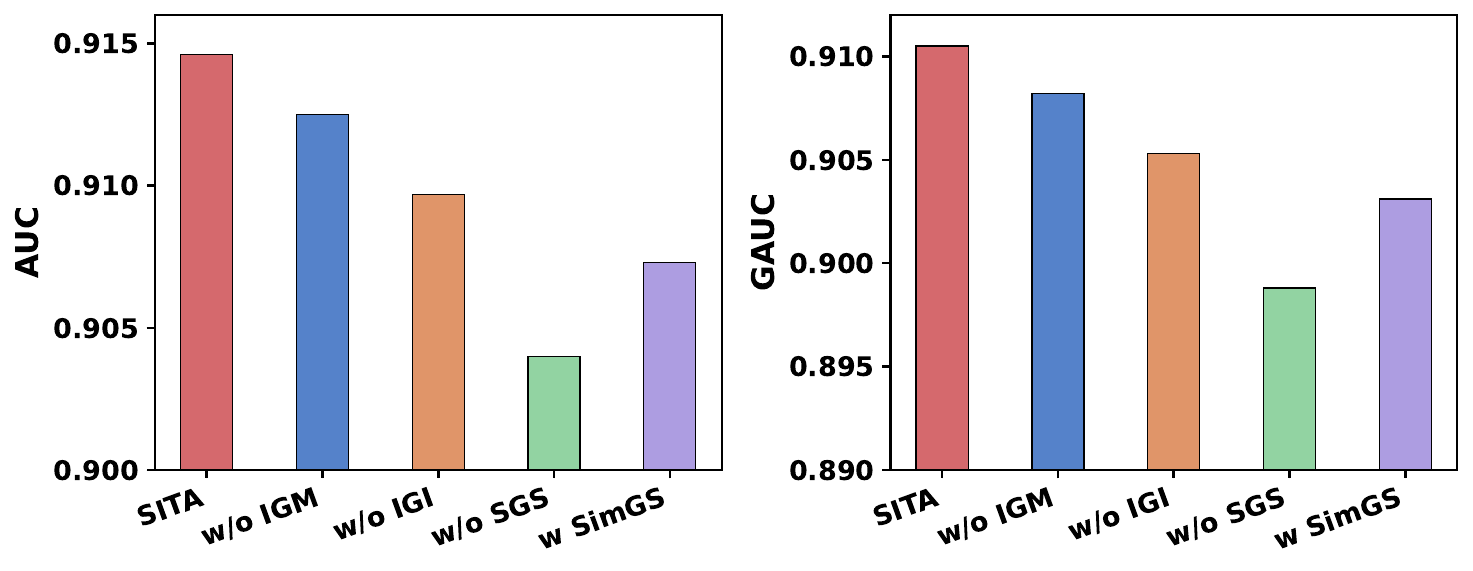}
  \caption{Ablation study of SITA. The left panel shows the AUC of SITA and its variants, while the right panel shows the corresponding GAUC.}
  \label{fig:ablation_study}
  \vspace{-3mm}
\end{figure}

Figure~\ref{fig:ablation_study} presents the ablation results of SITA. Removing any component consistently degrades the recommendation performance, demonstrating the effectiveness of each module in the proposed framework. We further compare three interest token selection strategies: SID-Guided Selection (SITA), Similarity-Guided Selection (SITA w/ SimGS), and without Selection (SITA w/o SGS), where all interest tokens are used. The results show that removing the selection mechanism leads to the largest performance drop, highlighting the importance of target-aware interest token selection. Although Similarity-Guided Selection partially recovers the performance by selecting target-related interest tokens, it still performs noticeably worse than SID-Guided Selection, indicating that the improvement comes not only from selecting different interest tokens for different target items, but also from the semantic correspondence between semantic identifiers and structured interest groups.

\subsection{Hyperparameter Study}
We further investigate the impact of two key hyperparameters in SITA: the semantic space configuration $(N,K)$ and the number of Structured Interest Compression (SIC) blocks $L$. Here, $N$ denotes the number of semantic groups and $K$ denotes the number of interest tokens within each group. 

\begin{figure}
  \centering
  \includegraphics[width=\linewidth]{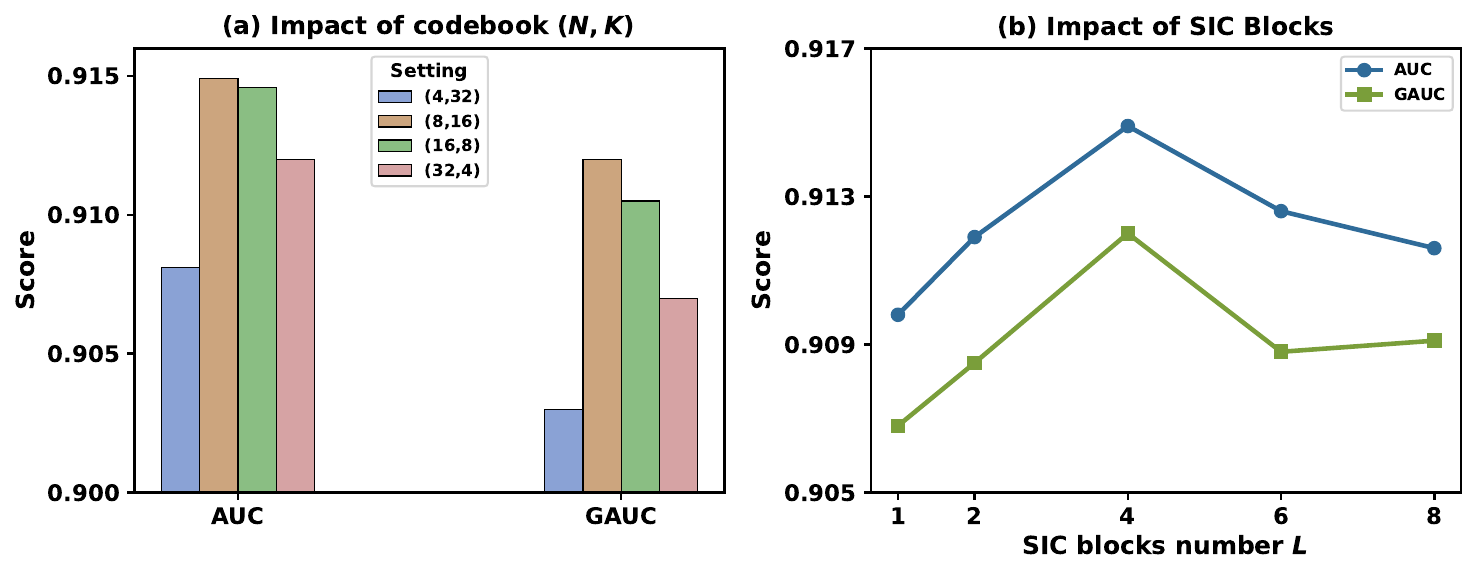}
  \caption{Hyperparameter analysis of SITA. The left subplot shows the effect of different semantic space configurations $(N,K)$, and the right subplot shows the effect of the number of SIC blocks $L$.}
  \label{fig:hyperparam_study}
  \vspace{-3mm}
\end{figure}

The results are presented in Figure~\ref{fig:hyperparam_study}. In the left subplot, we fix the total number of codewords $N \times K = 128$ while varying the semantic space configuration, which results in different numbers of representable patterns $K^N$. The balanced configuration achieves the best performance, suggesting that a proper balance between the number of semantic groups and the number of codewords within each group is beneficial for structured interest modeling. The right subplot shows that the performance first improves as the number of SIC blocks increases, reaches its optimum with four blocks, and then slightly declines, indicating that a moderate number of SIC blocks is sufficient to effectively model structured user interests.

\begin{figure}
  \centering
  \includegraphics[width=\linewidth]{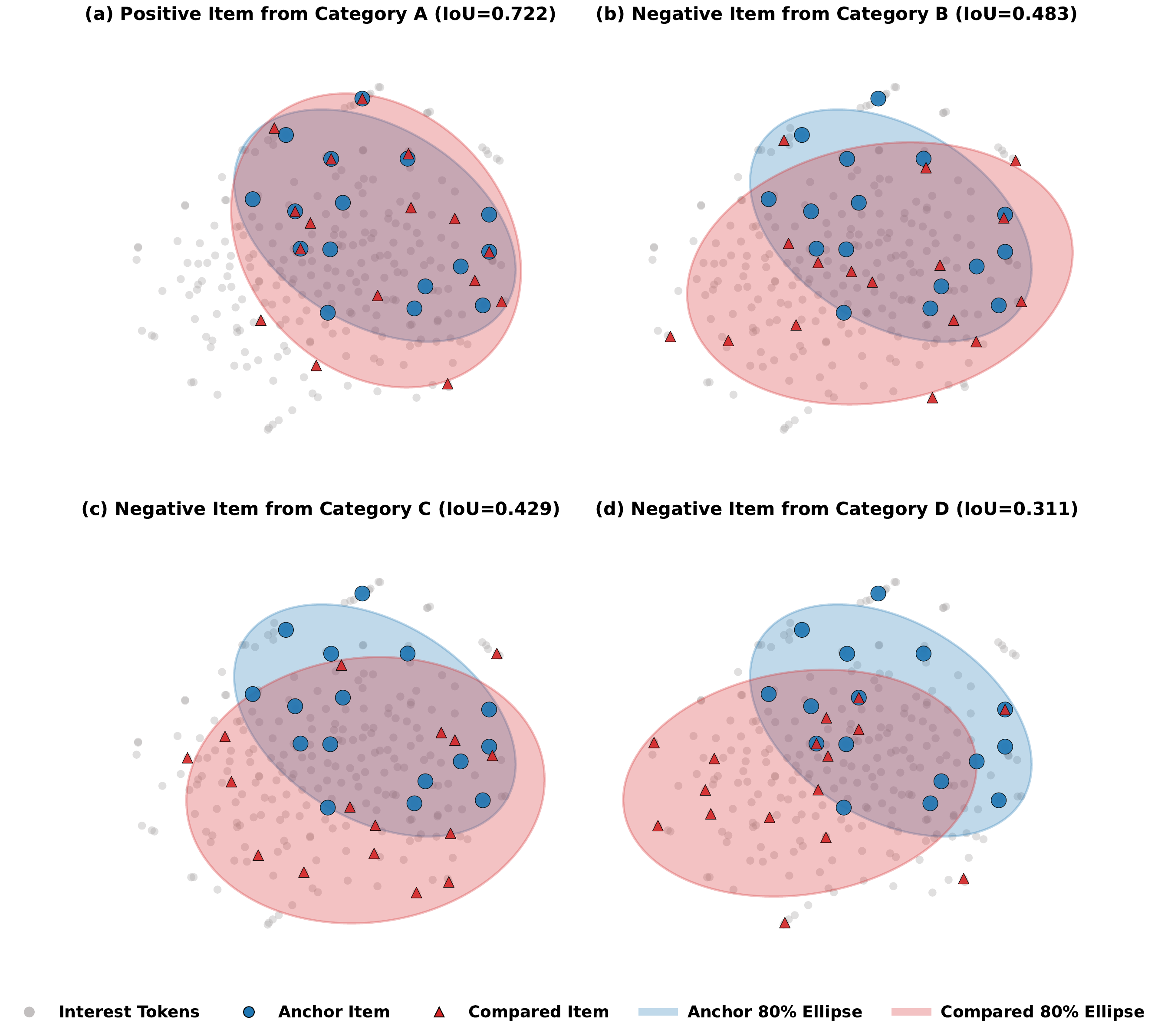}
  \caption{Visualization of target-aware selection in the learned interest-token space. Each subplot highlights the selected tokens of two items and their corresponding 80\% covariance ellipses, with IoU measuring the overlap between their regions.}
  \label{fig:case_study}
  \vspace{-2mm}
\end{figure}

\begin{figure*}
  \centering
  \includegraphics[width=0.95\linewidth]{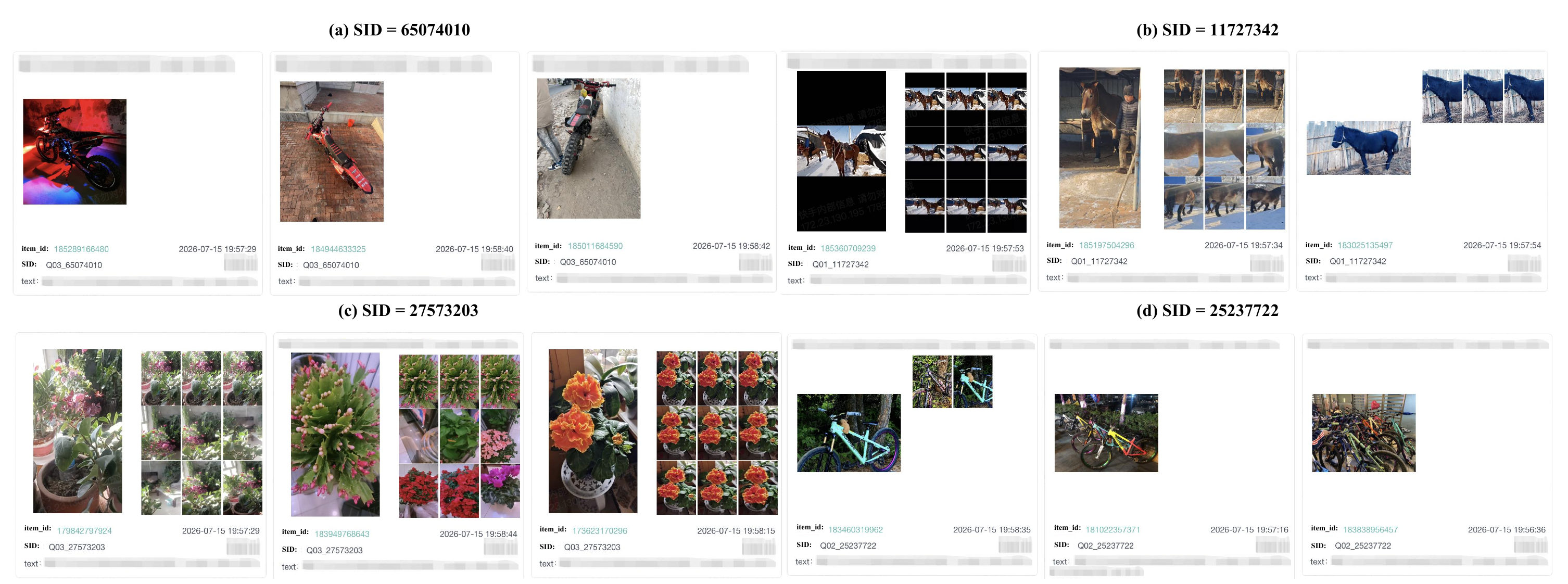}
  \caption{Examples of items sharing the same SID. Each row shows three representative items associated with one SID, where visually similar semantic patterns are consistently grouped under the same code, illustrating that the learned SID captures meaningful high-level semantics across items.}
  \label{fig:sid_analysis}
  \vspace{-2mm}
\end{figure*}
\subsection{Case Study}

To better understand how SITA leverages semantic identifiers for target-aware interest token selection, we conduct case studies from two perspectives: (1) whether SID-guided selection captures semantic relationships among target items, and (2) whether the learned semantic identifiers organize items with meaningful semantics.

\subsubsection{Target-Aware Interest Token Selection}

We first analyze the selected interest tokens on the Taobao-MM dataset. Specifically, we examine two properties: (1) \textbf{Inter-category separability}, where items from different categories select distinct interest regions; and (2) \textbf{Intra-category diversity}, where items from the same category select similar but non-identical interest tokens.

For an anchor item, we sample a positive item from the same category and a negative item from a different category. We visualize their selected interest tokens using t-SNE and characterize each selection region with an 80\% covariance ellipse. The similarity between two regions is measured by Intersection over Union (IoU).

Figure~\ref{fig:case_study} shows that same-category item pairs achieve higher IoU values than cross-category pairs, indicating that SITA selects semantically consistent interest tokens for similar items. Meanwhile, same-category items still preserve non-identical selection patterns, demonstrating fine-grained item-level diversity. These results verify that SITA enables target-aware interest token selection while maintaining a structured global interest space.

\subsubsection{Semantic Organization of Semantic Identifiers}

We further analyze whether the learned semantic identifiers capture meaningful item semantics on the industrial dataset. Specifically, we retrieve representative items sharing the same SID and examine their semantic consistency.

Figure~\ref{fig:sid_analysis} presents several examples of items with identical SIDs. For privacy protection, sensitive information is anonymized through mosaic masking. Items sharing the same SID exhibit strong semantic coherence and correspond to similar high-level concepts, such as motorcycles and flowers, demonstrating that BPQ effectively organizes items into a semantically structured space.

\subsection{Industrial Study}
To further validate the effectiveness of SITA in real-world recommender systems, we conduct experiments on a production recommendation platform serving hundreds of millions of daily active users. The online baseline compresses user behavior sequences into a set of shared user interests for long-sequence modeling. In the industrial evaluation, SITA replaces the original long-sequence modeling module while keeping the remaining recommendation framework unchanged. For fair comparison, SITA is initialized from the latest checkpoint of the online baseline, and newly introduced parameters are randomly initialized. Both models consume the same streaming training data throughout the evaluation period. The maximum behavior sequence length is set to 2,500.

Following the industrial evaluation protocol, we report the stabilized average AUC and GAUC of Effective-View under two production scenarios over a continuous evaluation window. Figure~\ref{tab:online_results} presents the relative improvements of SITA over the online baseline. Notably, given the stability brought by the massive scale of training samples, a relative improvement of approximately \textbf{0.05\%} is regarded as practically significant and can lead to substantial business impact in our production environment. SITA achieves consistent improvements across all reported metrics, validating that it successfully bridges target-aware modeling and global interest modeling under real-world serving constraints. By compressing the entire user behavior history into semantically structured interest tokens and enabling target-specific selection through semantic identifiers, SITA preserves global user interest modeling while improving target-aware interest matching, while the compact user representations maintain efficient online serving for industrial recommendation scenarios with extremely long behavior histories.



\begin{table}
  \centering
  \caption{Relative improvements in Effective-View across two industrial scenarios.}
  \label{tab:online_results}
  \setlength{\tabcolsep}{8pt}
  \renewcommand{\arraystretch}{0.95}
  \begin{tabular}{lcccc}
    \toprule
    \multirow{2}{*}{Model} & \multicolumn{2}{c}{Scenario 1} & \multicolumn{2}{c}{Scenario 2} \\
    \cmidrule(lr){2-3} \cmidrule(lr){4-5}
    & AUC & GAUC & AUC & GAUC \\
    \midrule
    SITA & \textbf{+0.057\%} & \textbf{+0.049\%} & \textbf{+0.078\%} & \textbf{+0.076\%} \\
    \bottomrule
  \end{tabular}
  \vspace{-3mm}
\end{table}

%% file: sections/conclusion.tex
\section{CONCLUSION}
In this paper, we revisited long-sequence recommendation from the perspective of target-aware global interest modeling. We identified that existing methods struggled to simultaneously achieve target-aware modeling, global interest modeling, and real-world deployability. To address this challenge, we proposed SITA, which introduced semantically organized interest tokens to obtain compressed representations of user behavior sequences and enabled candidate-specific selection for target-aware global interest modeling with efficient inference. Extensive experiments demonstrated that SITA successfully achieved target-aware global interest modeling while maintaining real-world deployability. We hope SITA will inspire future research on scalable, target-aware, and deployable long-sequence recommendation. Meanwhile, we will continue to explore further optimizations of SITA and investigate its broader deployment in real-world recommendation systems.